\documentclass[12pt]{article}
\usepackage{graphicx, url}
\usepackage{palatino}
\usepackage{amsmath, amssymb,bbm}
\usepackage{epsfig}
\usepackage{rotating}
\usepackage{dcolumn}
\usepackage{bm}
\usepackage{caption}
\usepackage{subcaption}
\usepackage[table]{xcolor}
\newcommand{\comment}[1]{}

\def \bea{\begin{eqnarray}}
\def \eea{\end{eqnarray}}

\def \qqbar{Q\Qbar}

\def \qqbar{q\overline{q}}

\begin{document}   
\baselineskip 18pt
\title{Bottomonia production in Modified NRQCD}
\author{
   Sudhansu~S.~Biswal$^1$\footnote{E-mail: sudhansu.biswal@gmail.com}, 
   ~Monalisa Mohanty$^1$\footnote{Email: monalimohanty97@gmail.com}
     ~and  K.~Sridhar$^2$\footnote{E-mail: sridhar.k@apu.edu.in} \\ [0.2cm]
    {\it \small 1. Department of Physics, Ravenshaw University,} \\ [-0.2cm]
    {\it \small Cuttack, 753003, India.}\\ [-0.2cm]
    {\it \small 2. School of Arts and Sciences, Azim Premji University,} \\ [-0.2cm]
    {\it \small Sarjapura, Bangalore, 562125, India.}\\
}
\date{}
\maketitle

\begin{abstract}
\noindent Motivated by the success of Modified Non-Relativistic Quantum Chromodynamics (Modified NRQCD) in explaining
data from experiments at the Large Hadron Collider (LHC) for charmonia, we now turn to the study of bottomonium 
production at the LHC. Modified NRQCD does very well in explaining $\Upsilon$ data from the LHC. But this is
true also of NRQCD which explains the $\Upsilon$ data equally well. Where the two models differ substantially is
in their predictions for $\eta_b$ production. As was the case with $\eta_c$, the measurement of $\eta_b$ production 
at the LHC will be another decisive test of Modified NRQCD. 
\end{abstract}

\maketitle

\noindent
There are few problems in Quantum Chromodynamics (QCD) where we can theoretically study the production of 
hadrons in high-energy collisions and these exceptional cases are that of the production of bound states of 
a heavy quark and an antiquark viz., quarkonia. 
The relative velocity $v$ of the heavy quark anti-quark pair is small, and so these bound states can be studied 
in an effective theory derived from QCD in the non-relativistic approximation called Non-Relativistic 
Quantum Chromodynamics (NRQCD) ~\cite{bbl}. One starts by neglecting, from the QCD Lagrangian, all states of 
momenta much larger than the heavy quarkonium mass, $M$, and accounting for this exclusion by adding new 
interaction terms yielding the effective Lagrangian. 

The quarkonium state admits of a Fock-state expansion in orders of $v$.  
In NRQCD at leading order, the  $Q\bar Q$ state is in a colour-singlet state, 
however at $O(v)$, it can be in a colour-octet state which is connected to the 
physical quarkonium state through non-perturbative gluon emissions.
The cross-section for the production of a quarkonium state $H$ of mass $M$ in NRQCD can be 
factorised as:
\bea
  \sigma(H)\;=\;\sum_{n=\{\alpha,S,L,J\}} {F_n\over {M}^{d_n-4}}
       \langle{\cal O}^H_n({}^{2S+1}L_J)\rangle, 
\label{factorizn}
\eea
where $F_n$'s are the short-distance coefficients and ${\cal O}_n$ are 
operators of naive dimension $d_n$, describing the 
long-distance effects. Due to NRQCD factorization, 
the non-perturbative matrix elements are energy independent 
and can be extracted at a given energy and
used in the 
prediction of quarkonium cross-sections at other energies.

NRQCD achieved much success in explaining 
charmonium production at the Fermilab 
Tevatron \cite{cdf}, in contrast to the 
colour-singlet model \cite{baier}. However, it does not 
predict the normalization of the $p_T$ distributions because 
of the unknown non-perturbative parameters that appear
in the NRQCD predictions. Independent tests of NRQCD are 
consequently important \cite{tests,bs1,brtn,tests2} and the 
prediction of polarisation of the produced quarkonium 
state is an important test. NRQCD predicts that at large
$p_T$ the quarkonium is transversely polarised but 
experiments see no evidence for this polarisation in
either $J/\psi$ \cite{jpsi_pol} or 
 $\Upsilon$ measurements \cite{CDF, CMS1, LHCb2}. 

Another problem, not completely unrelated to the 
polarisation puzzle, is the production of $\eta_c$. 
As in the case of polarisation, the heavy quark 
symmetry of the NRQCD Lagrangian is useful in predicting the $\eta_c$
cross-section \cite{bs1} but again the predictions are 
completely at variance with measurements from the LHCb 
experiment \cite{lhcb_2}.

In a recently proposed modification of NRQCD \cite{bms}, which we have named Modified NRQCD, we suggested that
the colour-octet $c \bar c$ state can radiate several
soft {\it perturbative} gluons -- each emission taking away little energy but carrying
away units of angular momentum. In the multiple emissions that the colour-octet state
can make before it affects the final NRQCD transition to a quarkonium state, the
angular momentum and spin assignments of the $c \bar c$ state changes constantly.
Consequently, the Fock expansion for $J/\psi$ (analogous to the one given in Eq.~\ref{fockexpn} for $\eta_c$)
is no longer valid in Modified NRQCD. 

In Refs.~\cite{bms} and \cite{bms2}, we have studied $J/\psi$ and $\chi_c$ production, respectively and
in Ref.~\cite{bms3} we have shown how Modified NRQCD can provide a solution to the $\eta_c$ anomaly. In this
paper we study $\Upsilon$ data from the LHC using Modified NRQCD. 

In NRQCD, the Fock space expansion of the physical $\Upsilon$, which is a $^3S_1$
($J^{PC}=1^{--}$) state, is: 
\bea
\left|\Upsilon\right>={\cal O}(1)
        \,\left|\qqbar[^3S_1^{[1]}] \right>+
         {\cal O}(v^2)\,\left|\qqbar[^3P_J^{[8]}]\,g \right>+ \cr
        {\cal O}(v^4)\,\left|\qqbar[^1S_0^{[8]}]\,g \right>+
        {\cal O}(v^4)\,\left|\qqbar[^3S_1^{[8]}]\,gg \right>+
           \cdots. 
\label{fockexpn}
\eea

In the above expansion the colour-singlet $^3S_1^{[1]}$ state 
contributes at $O(1)$. The $^3P_J$ contribution includes the 
three $^3P_J$ i.e. $^3P_0$, $^3P_1$ and $^3P_2$. As $P$ state 
production is itself down by factor of ${\cal O}(v^2)$ both the 
colour-octet  $P$ and $S$ channels effectively contribute at 
the same order. The colour-octet  $^3P_J^{[8]}$ ($^1S_0^{[8]}$) 
becomes the physical $\Upsilon$ by emitting a gluon in an E1(M1) transition, 
while there is also a contribution at the same order 
from a $^3S_1^{[8]}$ state doing double E1 transition to the 
$\Upsilon$ state . Using Eq. (\ref{factorizn}) the NRQCD cross-section formula 
when written down explicitly in terms of the octet 
and singlet states for the $\Upsilon$ is given as follows:

\begin{eqnarray}
\sigma_{\Upsilon}  = \hat F_{{}^{3}S_1^{[1]}} \times \langle {\cal O} ({}^{3}S_1^{[1]}) \rangle +
                \hat F_{{}^{3}S_1^{[8]}} \times \langle {\cal O} ({}^{3}S_1^{[8]}) \rangle +\cr
                 \hat F_{{}^{1}S_0^{[8]}} \times \langle {\cal O} ({}^{1}S_0^{[8]}) \rangle 
                + {1 \over M^2} \biggl\lbrack\hat F_{{}^{3}P_J^{[8]}} \times \langle {\cal O} 
                     ({}^{3}P_J^{[8]}) \rangle \biggr\rbrack .
\label{Fock}
\end{eqnarray}

But the corresponding formulae for Modified NRQCD change because of the emission of multiple soft gluons
that change the quantum numbers of the intermediate colour octet states. The details are provided in 
Ref.~\cite{bms} but the Modified NRQCD formula for the production of the $\Upsilon$ is

\begin{eqnarray}
\sigma_{\Upsilon} &=& \biggl\lbrack \hat F_{{}^{3}S_1^{[1]}} 
                \times \langle {\cal O} ({}^{3}S_1^{[1]}) \rangle \biggr\rbrack \cr 
                &+& \biggl\lbrack  
                  \hat F_{{}^{3}S_1^{[8]}} 
                 + \hat F_{{}^{1}P_1^{[8]}} 
                + \hat F_{{}^{1}S_0^{[8]}} + (\hat F_{{}^{3}P_J^{[8]}} ) \biggr\rbrack 
                \times ({\langle {\cal O} ({}^{3}S_1^{[1]}) \rangle \over 8}) \cr
                &+& \biggl\lbrack  
                  \hat F_{{}^{3}S_1^{[8]}} 
                 + \hat F_{{}^{1}P_1^{[8]}} 
                + \hat F_{{}^{1}S_0^{[8]}} + (\hat F_{{}^{3}P_J^{[8]}} ) \biggr\rbrack 
                \times \langle {\cal O}  \rangle ,
\end{eqnarray} 
 
where
\begin{equation}
     \langle {\cal O}  \rangle =
                \times \biggl\lbrack 
                 \langle {\cal O} ({}^{3}S_1^{[8]}) \rangle 
                + \langle {\cal O} ({}^{1}S_0^{[8]}) \rangle 
                + {\langle {\cal O} ({}^{3}P_J^{[8]}) \rangle \over M^2}
                    \biggr\rbrack . 
\end{equation}

\begin{figure}[h!]
\begin{center}
\includegraphics[width=15cm]{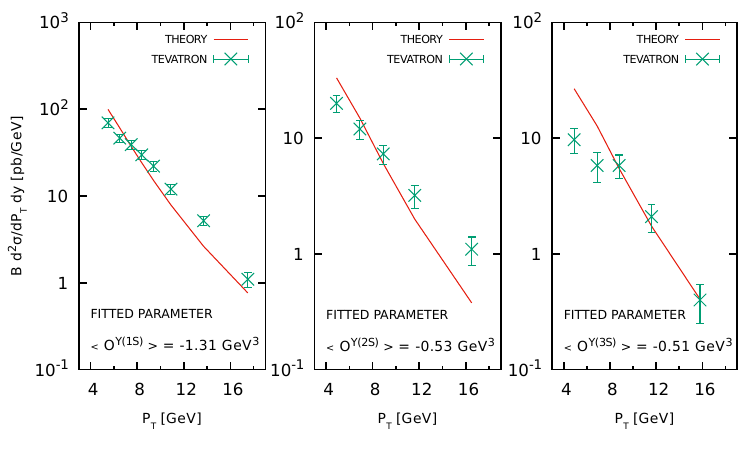}\caption{ Theoretical prediction of 
differential cross sections fitted to the data on $\Upsilon$(1S), 
 $\Upsilon$(2S) 
and $\Upsilon$(3S) production from the CDF experiment at Tevatron. }
        \label{fig:fig1}
\end{center}
\end{figure}

Thus instead of the three non-perturbative
parameters needed to get the $\Upsilon$ cross-section in NRQCD,
in this case it is just one unknown parameter ${\cal O}$ that needs to be obtained from the fit to the data.
We can fit this one parameter from using the data on the $\Upsilon\ (1S)$ 
states from Tevatron
and then use that to predict the $p_T$-distribution for this state and compare it with recent
$\Upsilon (1S)$ data from the LHC experiment. We do the same with the radial excitations: 
$\Upsilon (2S)$ and $\Upsilon (3S)$. 
The only other inputs we need to compute the $p_T$-distribution for $\Upsilon$ production are the subprocess matrix
elements and these have been calculated in Refs.~\cite{cho,gas,mat}. 

\begin{figure}[h!]
\begin{center}
\includegraphics[width=15cm]{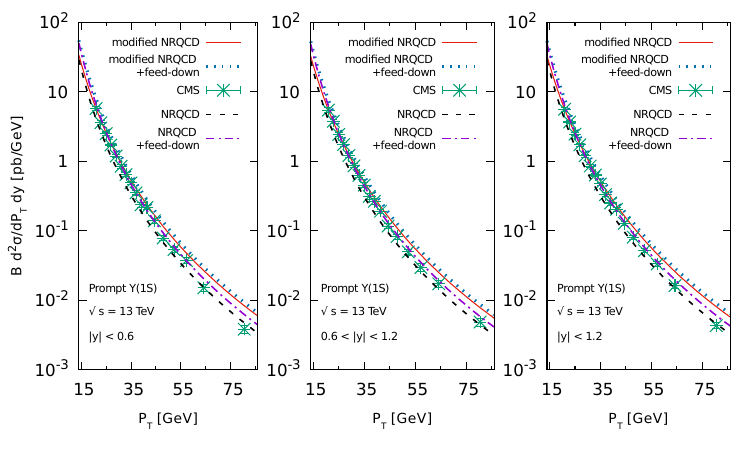}
\caption{Predicted double differential cross-sections for 
$\Upsilon$(1S) production at the LHC running at $\sqrt{s}$ = 13 TeV 
compared with the data collected from the CMS experiment.}
        \label{fig:fig2}
\end{center}
\end{figure}

\begin{figure}[h!]
\begin{center}
\includegraphics[width=15cm]{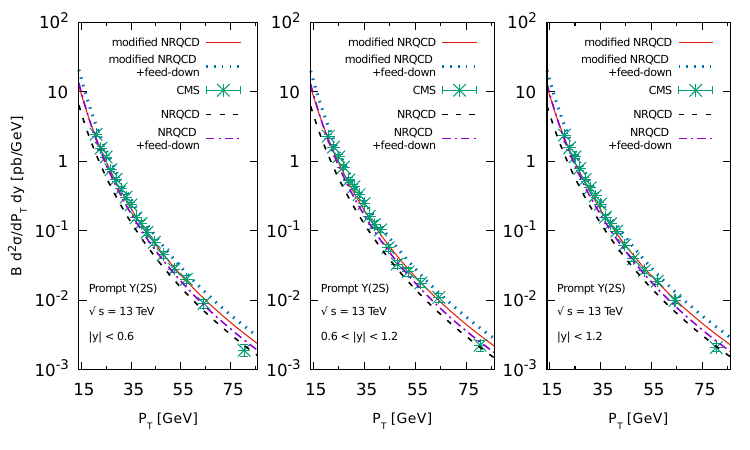}
\caption{Predicted double differential cross-sections for $\Upsilon$(2S) 
production at the LHC running at $\sqrt{s}$ = 13 TeV 
compared with the data collected from the CMS experiment.}
        \label{fig:fig3}
\end{center}
\end{figure} 

\begin{figure}[h!]
\begin{center}
\includegraphics[width=15cm]{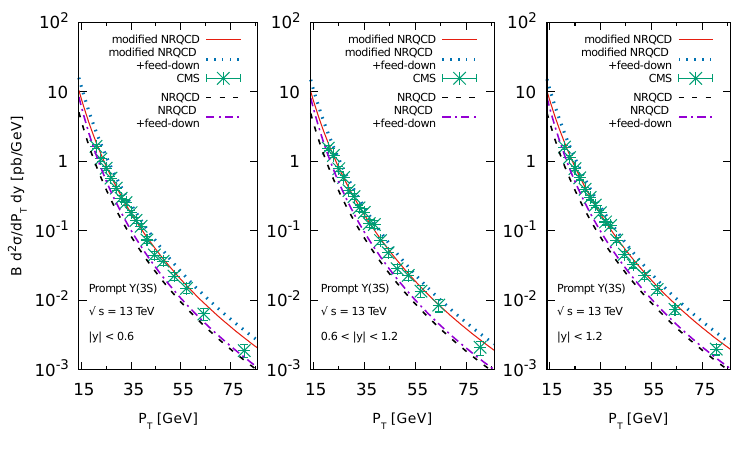}
\caption{Predicted double differential cross-sections for $\Upsilon$(3S) 
production at the LHC running at $\sqrt{s}$ = 13 TeV 
compared with the data collected from the CMS experiment.}
        \label{fig:fig4}
\end{center}
\end{figure}

The fits to the (somewhat sparse) Run 1 Tevatron data on $\Upsilon$ are shown in Fig.1.
But these fits are good enough to yield the value of the non-perturbative parameter ${\cal O}$
for the ${}^3S_1$ states: $\Upsilon (1S)$ and its radial excitations, the $2S$ and the $3S$ states.
We use these values of ${\cal O}$ to predict the $p_{T}$ distributions of these three states at the LHC 
and compare these with 13 TeV data from the CMS experiment \cite{CMS2}. 
These comparisons are shown in Figs. 2, 3 and 4 for the $1S$, $2S$ and $3S$ states, respectively. 
The predictions are in excellent agreement with the data 
from the CMS experiment in different rapidity ranges. Included in all three figures are the
predictions for the $S$-state with the feed-down from the $P$-state bottomonia also included. This
is done because the experiment does not distinguish direct production of the $S$-states from those
coming from the decay of the $P$-states. Our results show that the contribution of the $P$-states
via feed-down to the $S$ states is rather small. 

Also shown in Figs. 2, 3 and 4 are the NRQCD predictions for the $1S$, $2S$ and $3S$ states. As
for Modified NRQCD, for the NRQCD predictions we have done both the direct $S$-state production
and the $S$-state production including the feed-down from the $P$-states. For the NRQCD predictions
we have used the non-perturbative parameters given in Ref.~\cite{brtn}. 
The results
from NRQCD are very similar -- the agreement with the data is as good as in the Modified NRQCD case
and the contribution of the $P$-states is small. That there is not much to distinguish between the
two models in ${}^3S_1$ production is already a fact that we have noted in Refs. \cite{bms, bms2}
in the context of charmonia. 

\begin{figure}[h!]
\begin{center}
\includegraphics[width=15cm]{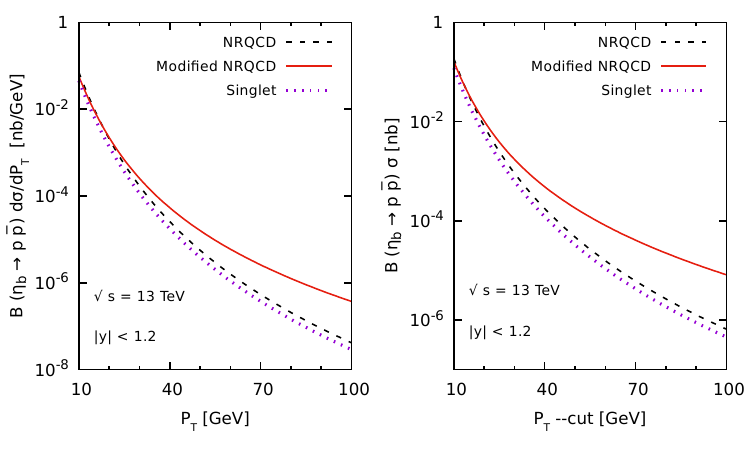}
\caption{Predictions for $\eta_{b}$ 
production at $\sqrt{s}$ = 13 TeV.}
        \label{fig:fig5}
\end{center}
\end{figure}

\begin{table}[h!]
\centering

\begin{tabular}{ |m{3cm}|m{3cm}|m{3cm}|m{3cm}| }
 \hline
\multicolumn{1}{|c|}{ } &\multicolumn{3}{c|}{$\sim$Expected number of events} \\[2mm]

\cline{2-4}
&&& \\
\multicolumn{1}{|c|}{Model} 
	&\multicolumn{1}{c|}{$P_{T}$ $>$ 20 GeV}&\multicolumn{1}{c|}{$P_{T}$ $>$ 30 GeV}&\multicolumn{1}{c|}{$P_{T}$ $>$ 40 GeV}
 \\[2mm]
 &&& \\

\hline
\hline
&&& \\
\multicolumn{1}{|c|}{NRQCD} &
	\multicolumn{1}{c|}{$1.62 
	\times 10^4$} &\multicolumn{1}{c|}{$1.83 
	\times 10^3$}&\multicolumn{ 1}{c|}{$3.53 
	\times 10^2$} 
 \\
 
&&& \\
\hline
&&& \\
\multicolumn{1}{|c|}{Modified NRQCD} &
	
	\multicolumn{1}{c|}{$2.05 
	\times 10^4$} &\multicolumn{1}{c|}{$3.47 
	\times 10^3$}&\multicolumn{ 1}{c|}{$9.72 
	\times 10^2$} 
\\
&&& \\
&&& \\
 \hline 
 
\end{tabular}
\caption{\label{tab:events}
Number of $\eta_b$ events expected at the LHC 
	running at $\sqrt{s}$ = 13 TeV for $\left|{\rm y}\right|~<~ 1.2$. 
}

\end{table}

However, if charmonia were to serve as a guide then we would expect to see strong model-dependence
in the study of the ${}^1 S_0$ state as was seen in the study of $\eta_c$ production in the two models 
\cite{bms3}. In the charmonium case, the LHC data on $\eta_c$ production strongly disagreed with the
predictions of NRQCD but was in very good agreement with those of Modified NRQCD. Motivated by these observations, 
we calculate the $\eta_b$ $p_T$ distributions in both Modified NRQCD and NRQCD. We study $\eta_b$ in its
decay into a $p \bar p$ state and have taken the branching ratio to be 
{\footnote { we have taken the branching ratio of  $\eta_c \to p \bar p$ 
	as the branching ratio of $\eta_b \to p \bar p$ 
is not available } }
 $1.35 \times 10^{-3}$ 
~\cite{pdg22}.

\begin{figure}[h!]
\begin{center}
\includegraphics[width=15cm]{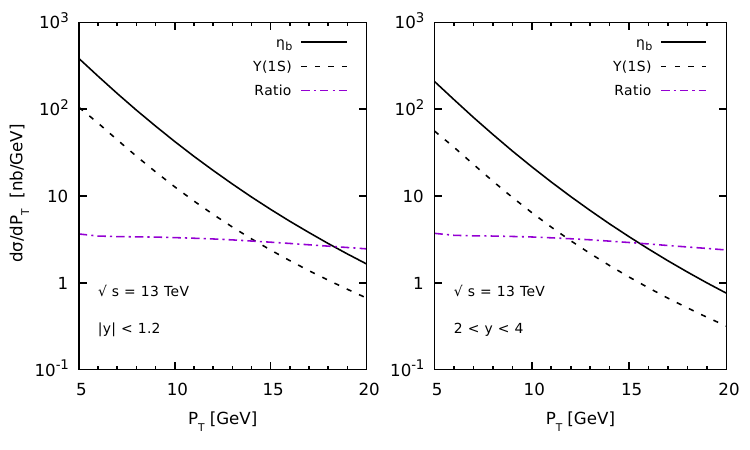}
\caption{Predicted ratio of $\eta_{b}$ to$\Upsilon$(1S) 
differential production cross-sections at $\sqrt{s}$ = 13 TeV using Modified NRQCD.}
        \label{fig:fig7}
\end{center}
\end{figure}

The results for $\eta_b$ production in the two models are presented in Fig. 5.  It can be seen that the Modified NRQCD 
 prediction is much larger than that of NRQCD. 
 Also shown in Fig. 5 is the colour-singlet
prediction. In contrast to the case of $\eta_c$, however, the NRQCD prediction for $\eta_b$ and the
colour-singlet prediction are very similar except maybe at very large $p_T$ and the Modified NRQCD prediction
is very different from both these predictions. 
A measurement of the $\eta_b$ $p_T$ distribution in LHC experiments can provide a crucial test of these
very interesting set of predictions. 

To get a sense of the feasibility of measuring the $\eta_b$ at the LHC, 
 we have also calculated the $p_T$-integrated 
cross-sections for different $p_T$ cuts, assuming an integrated luminosity
of $2\ {\rm fb}^{-1}$. These numbers are presented in Table 1 and suggest that one should expect a sizeable
number of $\eta_b$ events at the LHC experiments.

The ratio of the $\eta_b$ to $\Upsilon \ (1S)$ differential cross-sections is also a sensitive probe of
the mechanism of quarkonium production and we have presented in Fig. 6 the ratio of these two distributions
predicted by Modified NRQCD.

In conclusion, we have studied $\Upsilon \ (1S),\ (2S),\ (3S)$ production in Modified NRQCD
and NRQCD. The predictions for LHC agree well with data from the CMS experiment in both
models. As a model discriminating observable, we suggest the study of $\eta_b$ production
at the LHC. We show that there are huge differences in the $p_T$ distributions of $\eta_b$
in the two models. A measurement of this resonance and also a measurement of the ratio
of the $\eta_b$ to $\Upsilon\ (1S)$ cross-section should help 
 to discriminate between Modified
NRQCD and NRQCD and shed more light on the dynamics of quarkonium formation.

One of us (K.S.) gratefully acknowledges a research grant (No. 122500) from the Azim Premji
University. In this paper, as in everything else he writes, the results, opinions and views expressed are K.S.'s 
own and are not that of the Azim Premji University.



\end{document}